\documentclass[aps,superscriptaddress,prl,twocolumn]{revtex4-1}
\usepackage{graphicx}
\usepackage{epstopdf}
\usepackage{array}
\usepackage{color}
\usepackage{amsmath}
\usepackage{amsxtra}
\usepackage{amstext}
\usepackage{amssymb}
\usepackage{latexsym}
\usepackage{float}
\usepackage{color}
\usepackage[colorlinks=true, linkcolor=blue,urlcolor=blue,citecolor=blue]{hyperref}

\begin{document}

\title{High quality electrostatically defined hall bars in monolayer graphene}

\author{Rebeca Ribeiro-Palau}

\altaffiliation[Present address: ]{Centre de Nanosciences et de Nanotechnologies (C2N), CNRS, Univ Paris Sud,
Universit\'e Paris-Saclay, 91120 Palaiseau, France}
\email{rebeca.ribeiro@c2n.upsaclay.fr}
\affiliation{Department of Physics, Columbia University, New York, New York 10027, USA}
\affiliation{Department of Mechanical Engineering, Columbia University, New York,  New York 10027, USA}
\author{Shaowen Chen}
\thanks{S.Ch. and R. R.-P. contributed equally to this work}
\affiliation{Department of Physics, Columbia University, New York, New York 10027, USA}
\affiliation{Department of Applied Physics and Applied Mathematics, Columbia University, New York, New York 10027, USA}

\author{Yihang Zeng}
\affiliation{Department of Physics, Columbia University, New York, New York 10027, USA}
\author{Kenji Watanabe}
\affiliation{National Institute for Materials Science, 1-1 Namiki, Tsukuba 305-0044, Japan}
\author{Takashi Taniguchi}
\affiliation{National Institute for Materials Science, 1-1 Namiki, Tsukuba 305-0044, Japan}
\author{James Hone}
\affiliation{Department of Mechanical Engineering, Columbia University, New York, New York 10027, USA}
\author{Cory R. Dean}
\affiliation{Department of Physics, Columbia University, New York, New York 10027, USA}
\maketitle

{\bf Realizing graphene's promise as an atomically thin and tunable platform for fundamental studies and future applications in quantum transport requires the ability to electrostatically define the geometry of the structure and control the carrier concentration, without compromising the quality of the system. Here, we demonstrate the working principle of a new generation of high quality gate defined graphene samples, where the challenge of doing so in a gapless semiconductor is overcome by using the $\nu=0$ insulating state, which emerges at modest applied magnetic fields. In order to verify that the quality of our devices is not compromised by the presence of multiple gates  we  compare the electronic transport response of different sample geometries, paying close attention to fragile quantum states, such as the fractional quantum Hall (FQH) states, that are highly susceptible to disorder. The ability to define local depletion regions without compromising device quality establishes a new approach towards structuring graphene-based quantum transport devices.}

An important feature of two-dimensional (2D) electron systems is the ability to vary the charge carrier density by electrostatic gating. In semiconductor heterostructures this allows the geometry of the conducting region to be dynamically modified by using patterned gates to define local depletion regions. As a result, a variety of tunable device structures can be realized that allow the study and manipulation of 2D quantum transport phenomenon, ranging for example from single quantum point contacts\cite{vanHouten1996} to complex multi-terminal devices such as edge state interferometers\cite{vanWees1989,Chamon1997}. The recently developed graphene-based devices provide in principle a versatile new platform for the development of a new generation of quantum transport devices. The high quality of these samples is reflected by a carrier mobility which compares to the theoretical limit imposed by acoustic phonon scattering and a mean free path that can exceed the sample size \cite{Wang2013}, while the linear bandstructure and expanded degrees of freedom offer new capabilities beyond conventional systems\cite{Castro-Neto2009}. However, because monolayer graphene is gapless, it cannot be rendered insulating simply by depleting the region under the electrostatic gates.  While device structures can be shaped by lithographic patterning and etching, the resulting geometries are not tunable, and their response is typically dominated by the resulting edge disorder\cite{Bischoff2015}. Previous attempts to electrostatically define a channel in monolayer graphene, in the quantum Hall regime, lacked of an insulating state, complicating the control on edge states and understanding of the system \cite{Wei2017,Zimmermann2017,Kim2016,Nakaharai2011}.

  \begin{figure*}[t]
\centering
\includegraphics[scale=0.65]{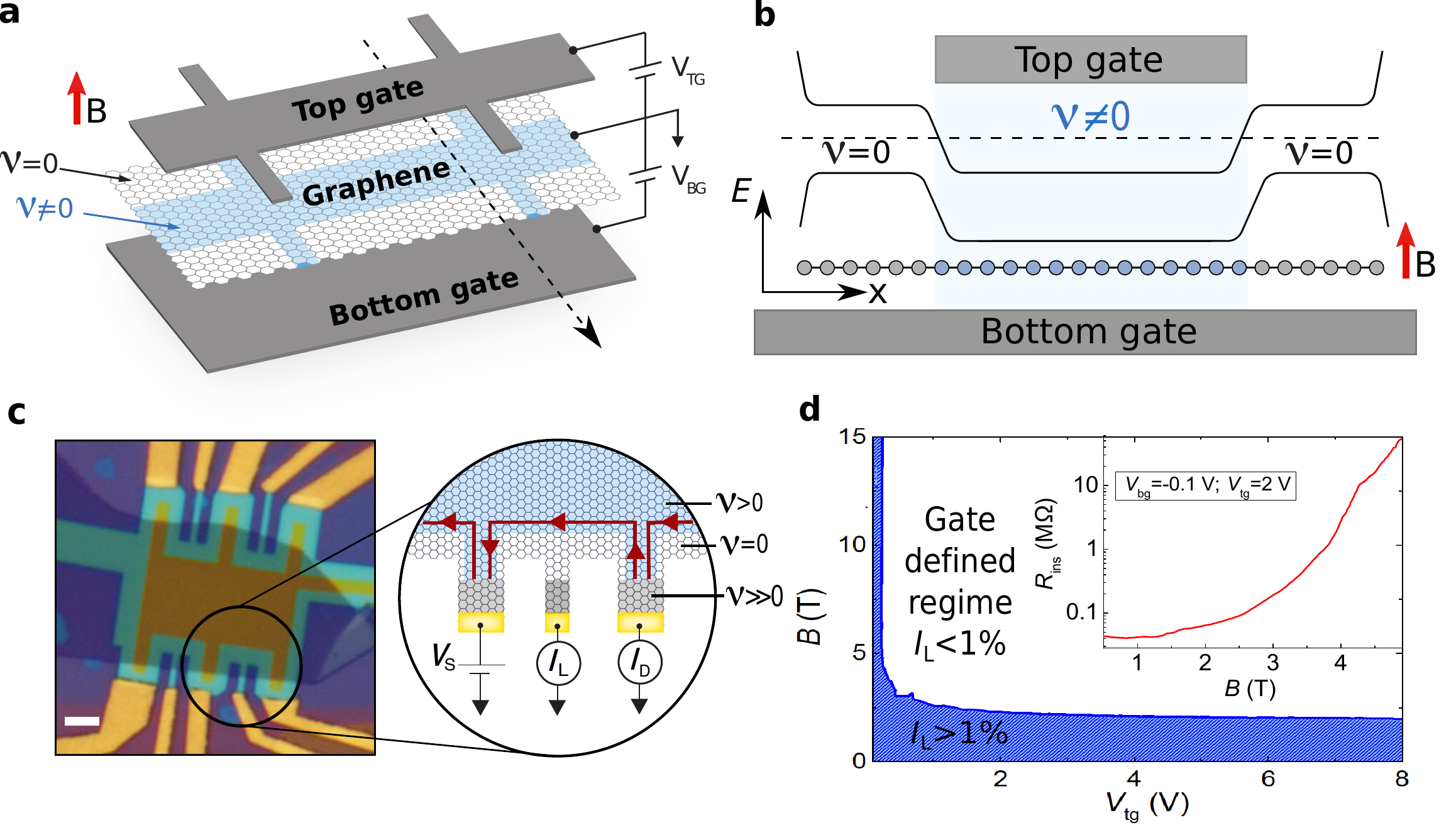}  
\caption{{\bf Electrostatically defined Hall bar in monolayer graphene}. {\bf a}, Schematic of the device and {\bf b}, cross-section and energy dispersion. In {\bf a} and {\bf b} the dielectric layer (BN) have been omitted for clarity. The graphite bottom-gate is used to set the Fermi energy of the external (uncovered) part of the device in the $\nu=0$ energy gap while the top-gate is use to vary the Fermi energy of the device. {\bf c}, Optical image of the device, scale bar 3 $\mu$m. Insert, schematic of the two types of contacts: i) regular source/drain/voltage contacts where the top gate area (blue) covers up to the highly doped graphene (gray) and ii) leak current leads used to probe if the external  part of the device is in an insulating state. {\bf d}, Leak current as a function of magnetic field and top gate voltage ($T=0.3$ K), blue area represent the region where the leak current is  $I_{\rm L}\geqslant1\%$ of the applied current. Insert: magnetic field sweep of the two probes resistance of the uncovered region for a fix top and bottom-gate voltage. }
\label{Device}
\end{figure*}

Here we demonstrate the working principle of a new generation of high--quality gate--defined monolayer graphene devices where an insulating state underneath the gate is achieved using the properties of graphene under magnetic field. To confirm the high quality of our samples we compare the  electronic transport response of fragile quantum states, such as the fractional quantum Hall effect (FQHE) states,  in different sample geometries. These states are used as sensitive indicators of quality. In a conventional 2D electron gas (2DEG) the use of electrostatic gates to change the carrier density and/or to define electrostatically is generally found to compromise the quality of the 2DEG \cite{Pan2017,Bachsoliani2017}. We observe a similar result in graphene when using evaporated metal gates but find that exfoliated graphite gates allows us to maintain high mobility.

Fig.  \ref{Device}a-c illustrates the working principle in our devices.  In a  sufficiently large perpendicular magnetic field the N=0 Landau level (LL) in graphene splits into sub-levels with an antiferromagnetic state appearing at $\nu=0$ that is characterized by being gapped both in the bulk and at the edges\cite{Young2012,Young2014, Bolotin2009,Amet2014}. Using the bottom gate we tune the entire device into this $\nu=0$ state. We then apply a finite bias to a patterned top gate, defining the active region. Both the top and bottom gate are separated from the graphene channel by multi-layer hexagonal boron nitride (BN) (not shown in Fig. 1a). To fabricate these devices we  assemble the heterostructure using the van der Waals assembly technique\cite{Wang2013}.  We then use two successive etching steps first to shape the entire structure into a multi-terminal Hall bar, then a second etch to further shape the top gate to a smaller Hall bar.  An optical image of the final device is shown in Fig. 1c.  A portion of the graphene leads extends past the bottom graphite gate before making contact with the evaporated edge-contact\cite{Wang2013}.  In all of the measurements presented here the carrier density of this extended lead region is tuned to high density (using the Si bottom gate) to  ensure good electrical contact in the QHE regime\cite{Maher2014} (see supplementary information).

\begin{figure*}[t!]
\centering
\includegraphics[scale=0.6]{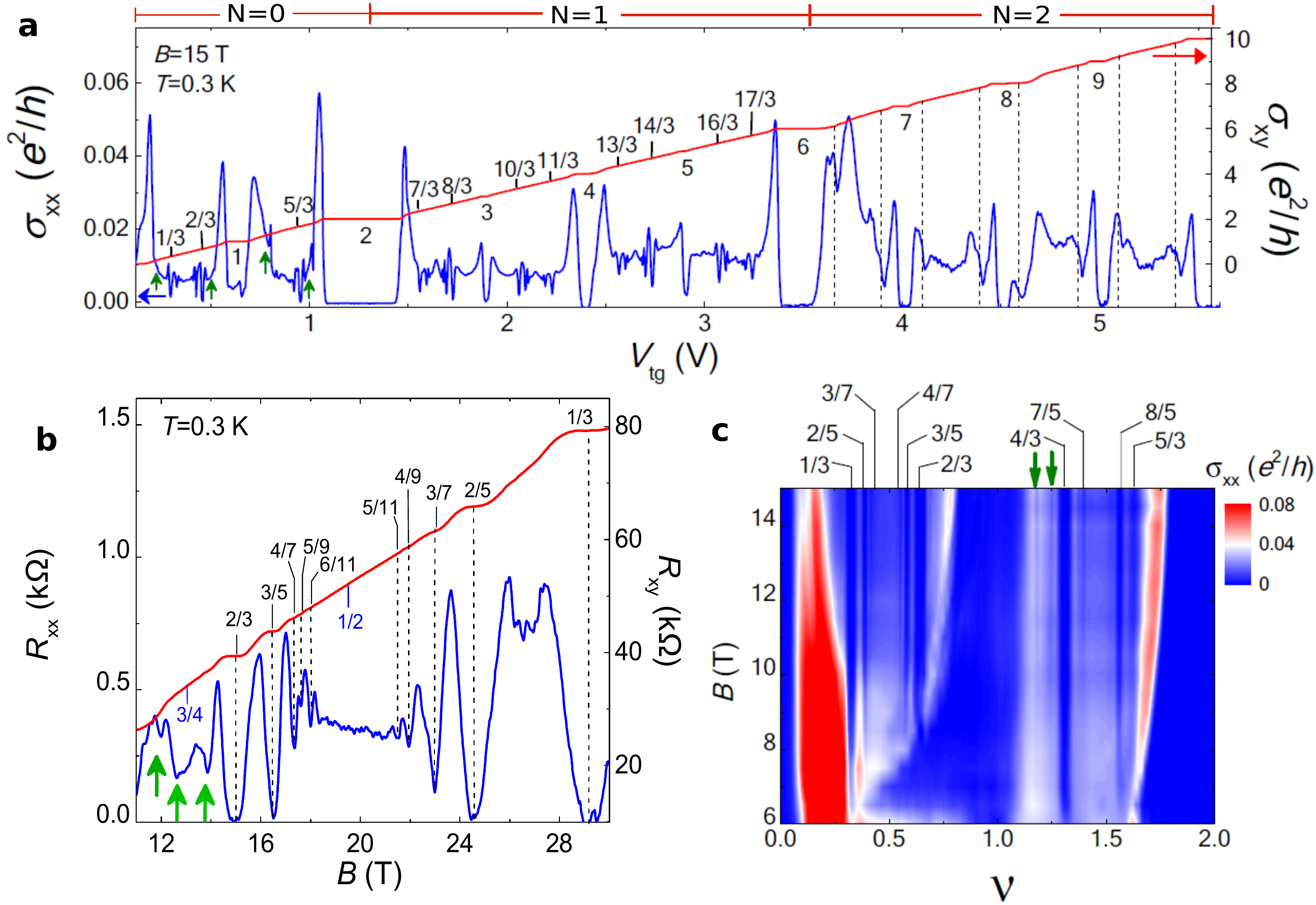}  
\caption{{\bf Improved FQHE in graphite gate defined monolayer graphene}. {\bf a}, Longitudinal (left) and Hall conductivity (right)  in the gate defined regimen at 15 T, only multiples of $1/3$ states have been labeled for clarity. {\bf b}, Magnetic field sweep  of the N=0 LL in the gate defined regime. {\bf c},  Longitudinal conductivity in the N=0 LL as a function of magnetic field and filling factor in the gate defined regime for a second sample, at 0.3 K. In all figures green arrows show four-flux CF states.}
\label{15T}
\end{figure*}

In order to confirm the insulating behaviour in the  single-gated depletion regions, the devices also incorporate extra leads (thinner electrodes seen in Fig. \ref{Device}c) that are used to measure any leakage current $I_{\rm L}$ (Fig. \ref{Device}d) through the insulating region.  As a benchmark we consider the device to be in the {\it gate defined regime} whenever the leakage current  measures  less than $1$\%  of the total  current flowing through the device. The combination of magnetic field and gate voltage required to achieve this state is shown by the white area in Fig. \ref{Device}d. An almost exponential increase of the resistance of the insulating state $R_{\rm{ins}}$ as a function of the magnetic field can be observed in the insert of Fig. \ref{Device}d.

Fig. \ref{15T}a shows the longitudinal ($\sigma_{\rm xx}$) and Hall ($\sigma_{\rm xy}$) conductivity at $B=15$ T as a function of the top-gate voltage for a device operated in the {\it gate defined regime}. The result shows very high quality transport response with  a large number of FQHE states observable in the first two LLs (N=0 and N=1). In addition, onset of multiples of $1/5$ states can be observed in the longitudinal conductivity for the N=2 LL. Similarly to the fractional  states previously reported in the third LL of ultra-high mobility GaAs/AlGaAs samples \cite{gervais2004competition}. Figure \ref{15T}b shows the longitudinal and Hall resistance for the N=0 LL versus magnetic field measured up to $B=30$ T.  In addition to the sequence of two-flux composite fermion (CF) states\cite{jain2007} around $\nu=1/2$, we observe the presence of four-flux CF states around $\nu=3/4$. These $^{4}$CF states have previously only been observed  in local electronic compressibility measurements in suspended graphene  \cite{Feldman2013, Feldman2012}. 

Fig. \ref{15T}c shows evolution of the $N=0$ FQHE states with magnetic field. The strongest $n/3$ states persist to less than 6~T.  This is among the lowest fields at which  transport signatures of the FQHE have been observed in graphene Hall bars, further confirming the high quality of our device\cite{Amet2015}.

\begin{figure*}[t!]
\centering
\includegraphics[scale=0.6]{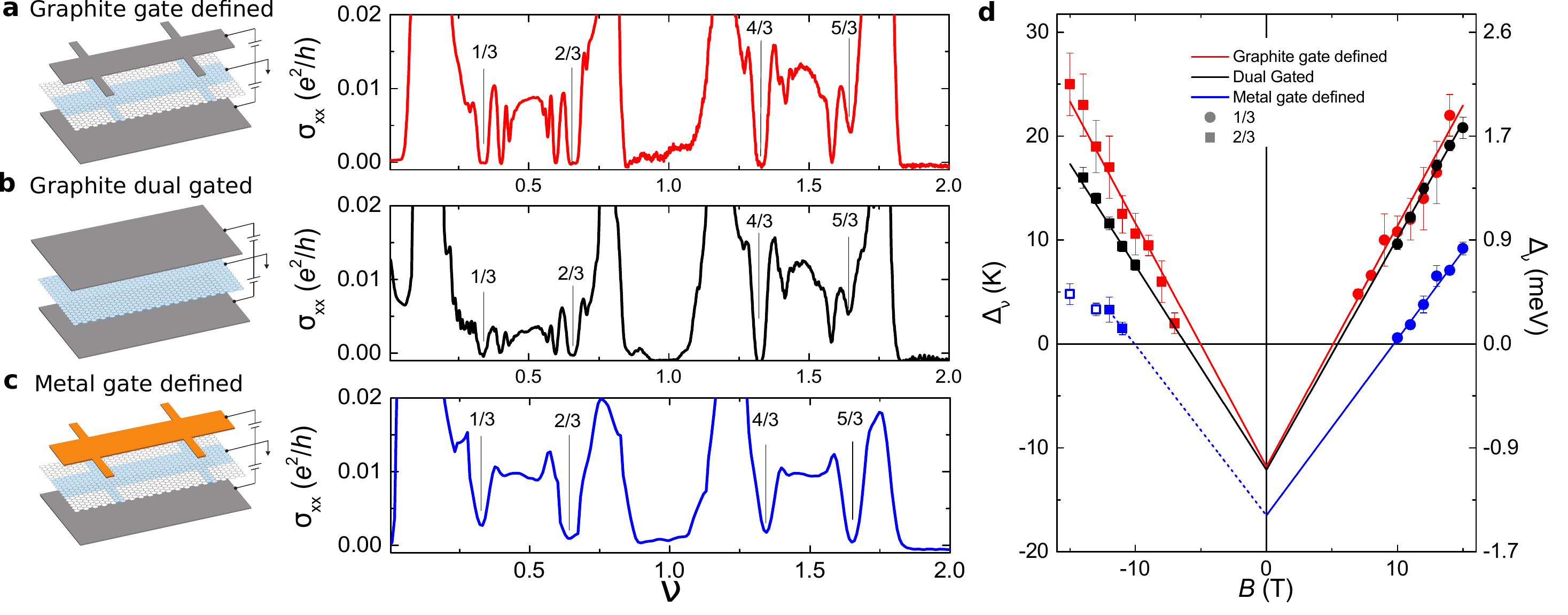}  
\caption{{\bf FQH states for different device geometries}.  Longitudinal  conductivity, $\sigma_{\rm xx}$, as a function of filling factor at $B=15$ T and $T=0.3$ K  for:  {\bf a}, graphite gate defined,  {\bf b}, graphite dual gated devices  and {\bf c}, metal gate defined. {\bf d}, Energy gaps  for $1/3$ and $2/3$ fractional quantum Hall states for a graphite gate defined (red), graphite  dual gated - second device - (black)  and metal gate defined (blue) as a function of magnetic field, values of the 1/3 state have been plotted in negative field for clarity. Solid lines represent linear fit. Dashed blue line represents linear fit of the two lower points of the 1/3 state for the metal gate defined, the other two measurements (empty symbols) seem to be compromised by the proximity of the $\nu=0$ insulating state.}
\label{gaps}
\end{figure*}

In our device structure the active region is fully encapsulated in graphite gates, which is a geometry that previously has been shown to significantly improve transport response.  In an effort to understand what role this plays in our device response we compare measurements from three device geometries, where in all cases we maintain a global bottom graphite gate, and vary details of the top gate.  The structures we considered (illustrated in Fig. 3a-c) include (i) a patterned graphite top gate (as described above) operated in the gate-defined regime, (ii) a global graphite top gate in which the device edges are defined by lithographic etching and (iii) a similar gate-defined device as in (i), but with an evaporated metal forming the top gate.

We first observe that, compared to the metal-gated structure, both of the graphite-gated devices show a larger number of well developed FQH states. This suggests that the use of a graphite gate is playing a role in improving the quality of the device and that, as in the case of conventional 2DEGs, the use of metal gates adversely affects device quality. A more quantitative comparison among the devices of Fig. \ref{gaps}a-c is provided by measuring the magnetic field dependence of the energy gaps of the  FQH states, $\Delta_{\nu}$ (Fig. \ref{gaps}d).  These energy gaps are extracted by thermal activation measurements, $R_{\rm xx}(T)\propto\exp(-\Delta_{\nu}/2kT)$, where $k$ is Boltzmann's constant. We assume that the  1/3 and 2/3 FQH gaps, which follow approximately linear in B dependence, correspond to the CF Zeeman energy  $\Delta_{\nu}=\frac{1}{2}\mu_{\rm B}gB+ \Gamma_{\nu}$, where $\mu_{\rm B}$ is the Bohr magneton and $g$ is the Land\'e g-factor \cite{Dethlefsen2006,Zeng2018,Schulze-Wischeler2004}. The intercept of this linear fit, gives the  broadening of the CF Landau levels (or $\Lambda$-levels), $\Gamma_{\nu}$, providing a quantitative comparison of sample disorder. The value of $\Gamma$ for the graphite gate defined and dual graphite gated devices is almost identical $\Gamma_{\nu}\approx 12$ K, and also in excellent agreement with recent corbino measurements \cite{Zeng2018}. We note that the two graphite gated devices fabricated from the same heterostructure however a similar response was found in another similarly constructed sample (see SI). In contrast, the metal gate-defined device clearly shows a larger $\Gamma_{\nu}$ than the two graphite-gated devices. This suggests that the use of two graphite gates results in a lower bulk disorder compared with evaporated metal gates,  consistent with other recent studies \cite{Zibrov2016}. The origin of the increased disorder in devices with metal gates  remains to be explored.  
 
High quality  FQHE in graphene at lower magnetic fields is an important achievement by itself since it enables the study of CF in a more tunable material. For example, using the same analysis we also extracted a Land\'e g-factor ranging between 6.9 and 4.9, which suggests  strong electronic interaction and possible spin textures, already proposed to exist in graphene\cite{Zeng2018,Dean2011}. Additionally, these enhanced electronic interactions in our sample are observed by the presence of a reentrant integer quantum Hall effect (RIQHE) at higher magnetic fields, as reported in \cite{Chen2018}. 

Finally we consider the nature of the QHE edge state in these gate defined Hall bars.  Due to the close proximity and sharp termination of the patterned graphite gate, the confinement potential may be substantially different from bottom-gated devices with etched boundaries\cite{Cui2016,Weis2011}.  In the gate defined and dual gated  cases we expect the new electrostatic profile to be soft (varying over 60 nm for our devices, see supplementary information). A broader confinement will be reflected in a larger spacial separation between edge states which could have a significant impact on details of the FQHE edge transport\cite{Grivnin2014}.  While detailed consideration of these effects is beyond the present work, the ability to modify the electrostatic profiles by choice of BN dielectric thickness provides an opportunity explore these effects in future experiments.

To summarize, we demonstrate that electrostatic gating can be used to define the geometry of graphene devices by utilizing the $\nu=0$ gap of the quantum Hall effect to maintain an insulating state. Whereas metal gates introduce disorder, graphite gates are compatible with ultrahigh-quality devices, as assessed by measurement of FQH response and observation of electron solid states through observation of RIQHE. 
The ability to define local depletion regions without compromising device quality establishes a new approach towards structuring graphene-based quantum transport devices. In particular our results establish the capability to combine robust FQHE states with complex  structures in graphene such as quantum point contacts and edge state interferometers. These are the essential pieces for the possible study of fractional and non-Abelian statistics. 

\bibliography{references}

\section*{Acknowledgements}

 We acknowledge discussions with M. O. Goerbig. S. Chen is supported by the ARO under MURI W911NF-17-1-0323.  This research was supported by the NSF MRSEC programme through Columbia in the Center for Precision Assembly of Superstratic and Superatomic Solids (DMR-1420634). A portion of this work was performed at the National High Magnetic Field Laboratory, which is supported by National Science Foundation Cooperative Agreement No. DMR-1157490 and the State of Florida.
 
 \section*{Author Contributions }
 
R.R.-P., S.C. and C.R.D. designed the experiment. S.C. and Y.Z. fabricated the samples. R.R.-P. and S.C performed the experiments, analyzed the data and wrote the paper. T.T. and K.W. grew the crystals of hexagonal boron nitride. J.H. and C.R.D. advised on experiments, data analysis and writing the paper.

\clearpage


\newcommand{\bq}{{\bf q}}
\newcommand{\bp}{{\bf p}}
\newcommand{\br}{{\bf r}}
\newcommand{\bR}{{\bf R}}
\newcommand{\rhobar}{\bar{\rho}}
\newcommand{\nubar}{\bar{\nu}}

\renewcommand{\thefigure}{S\arabic{figure}}
\renewcommand{\thesubsection}{S\arabic{subsection}}
\renewcommand{\theequation}{S\arabic{equation}}
\setcounter{figure}{0} 
\setcounter{equation}{0}

\section*{Supplementary Information}

\section{Si gated contacts}

\begin{figure}[h]
\centering
\includegraphics[scale=0.4]{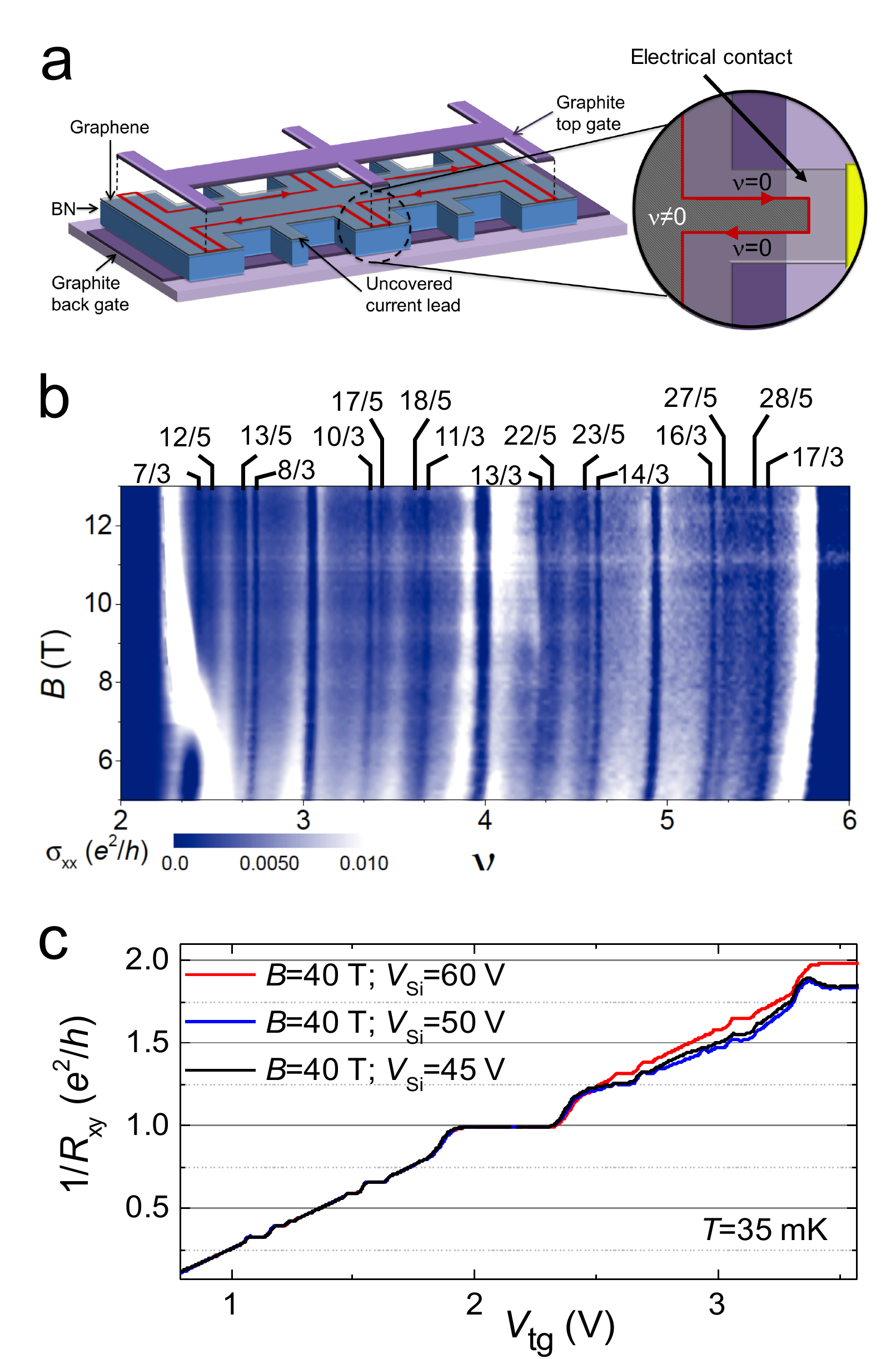}  
\caption{{\bf Si gated contacts}. {\bf a}, Device schematics with zoom in the contacts, top BN has been omitted for simplicity. {\bf b}, longitudinal conductance as a function of filling factor and magnetic field in the N=1 LL of a graphite gate defined sample at 0.3 K. {\bf c}, Resistance as a function of top gate voltage at 40 T and 35 mK for different Si gate values.}
\label{contacts}
\end{figure}

Electrical contacts play an important role in device performance. As explained in the main text, in our devices a portion of the graphene leads extends out of the graphite bottom local gate. This region is used as a tunable electrical contact controlled by the global Si back gate, see Fig. \ref{contacts}a. In all our experiments the Si gate is set inside a high index Landau level (corresponding to a $V_g>>60$ V) since the separation between LL decreases as the doping increases making  easier to keep the graphene at the leads in a metallic state.

Effects of an inadequate doping of the contacts can be seen for example in the 2D map of filling factor versus magnetic field of a graphite gate-defined device, as the one presented in Fig. \ref{contacts}b. In this,  the horizontal white areas, which correspond to an artificial increase of the longitudinal conductivity, are an effect of the LLs developing at the contacts as evidence by their periodicity in 1/$B$. In Fig. \ref{contacts}c  we can also see the impact of changing the Fermi energy of the contacts. As the Fermi energy  of the Si gated contacts is moved from the inside of a LL it passes from a metallic to a quantum Hall regime where the edge states of the Hall bar and those developed at the contact will interact generating a deviation from the quantized values.  \\

\begin{figure}[h]
\centering
\includegraphics[scale=0.3]{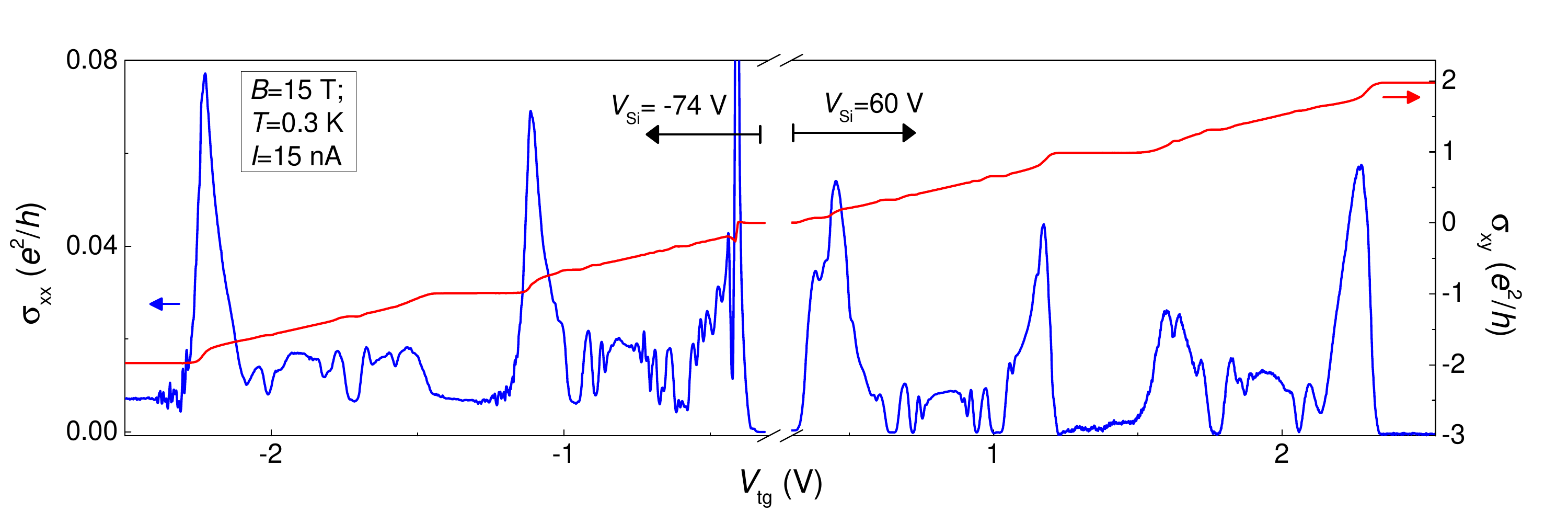}  
\caption{{\bf Si gated contacts, electron-hole symmetry}. Longitudinal and hole conductance for a graphite gate defined sample (GDHB-B) at 0.3 K and 15 T, the polarity of the Si gate voltage has been inverted in order to access the hole doped regime.}
\label{ElectronHole}
\end{figure}

Using these doped graphene contacts it is necessary to change the nature of the doping carriers by inverting the polarity of the Si gate voltage in order to access the hole regimen, Fig. \ref{ElectronHole}. It is important to mention that even when the hole doped regime can be access this one shows a lower quality than the electron doped regime.

\section{Gate-defined devices characterization}

\begin{figure}[h]
\centering
\includegraphics[scale=0.45]{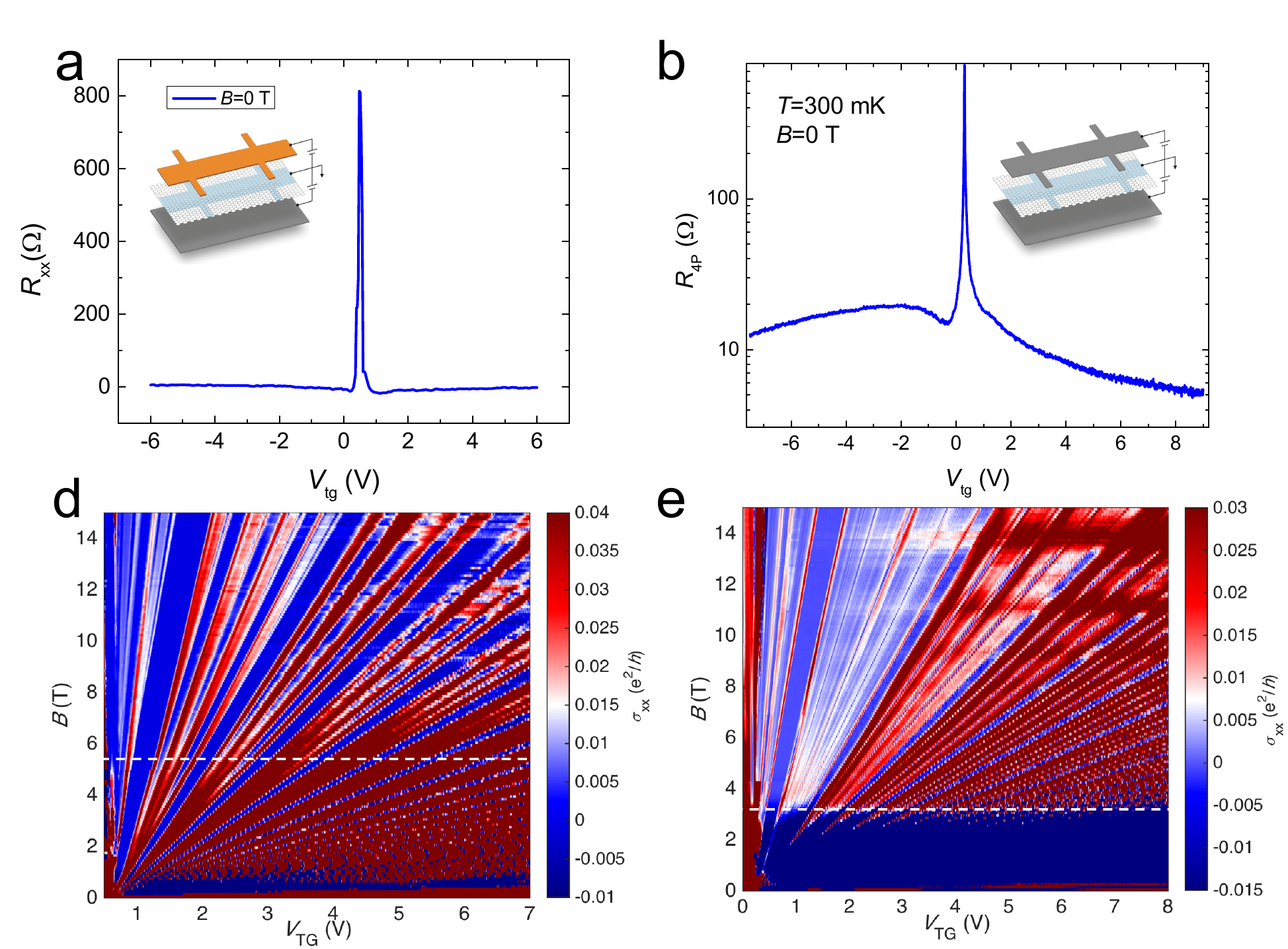}  
\caption{{\bf Devices characterization}. {\bf a} and {\bf b}, zero magnetic field transport for  metal and graphite gate defined, respectively. {\bf c} and {\bf d} Magnetic field versus top gate voltage for the same devices. Dashed line represents where the gate defined regime is achieved $I_{\rm leak}/I_{\rm total}<1\%$.}
\label{characterization}
\end{figure}

The zero field response of the three devices show in Fig. \ref{characterization}a-c was taken with the bottom gate at the CNP and sweeping the top gate. None of these devices show characteristics of BN/graphene alignment: satellite peaks or insulating state at the CNP. The extracted mobilities are all $\approx 200 000$ cm$^2$V$^{-1}$s$^{-1}$.

In Fig. \ref{characterization}d-f the gate and magnetic field sweep for two devices, from these we can see that the magnetic field at which  gate-defined regime achieved, as expected, also dependent on the sample disorder. In the main text we show that the metal gate defined posses a larger bulk disorder, this can also be seen in the need of a larger magnetic field to achieve the gate-defined regime or in other words to fully open the $\nu=0$ gap.  

\begin{figure}[h]
\centering
\includegraphics[scale=0.4]{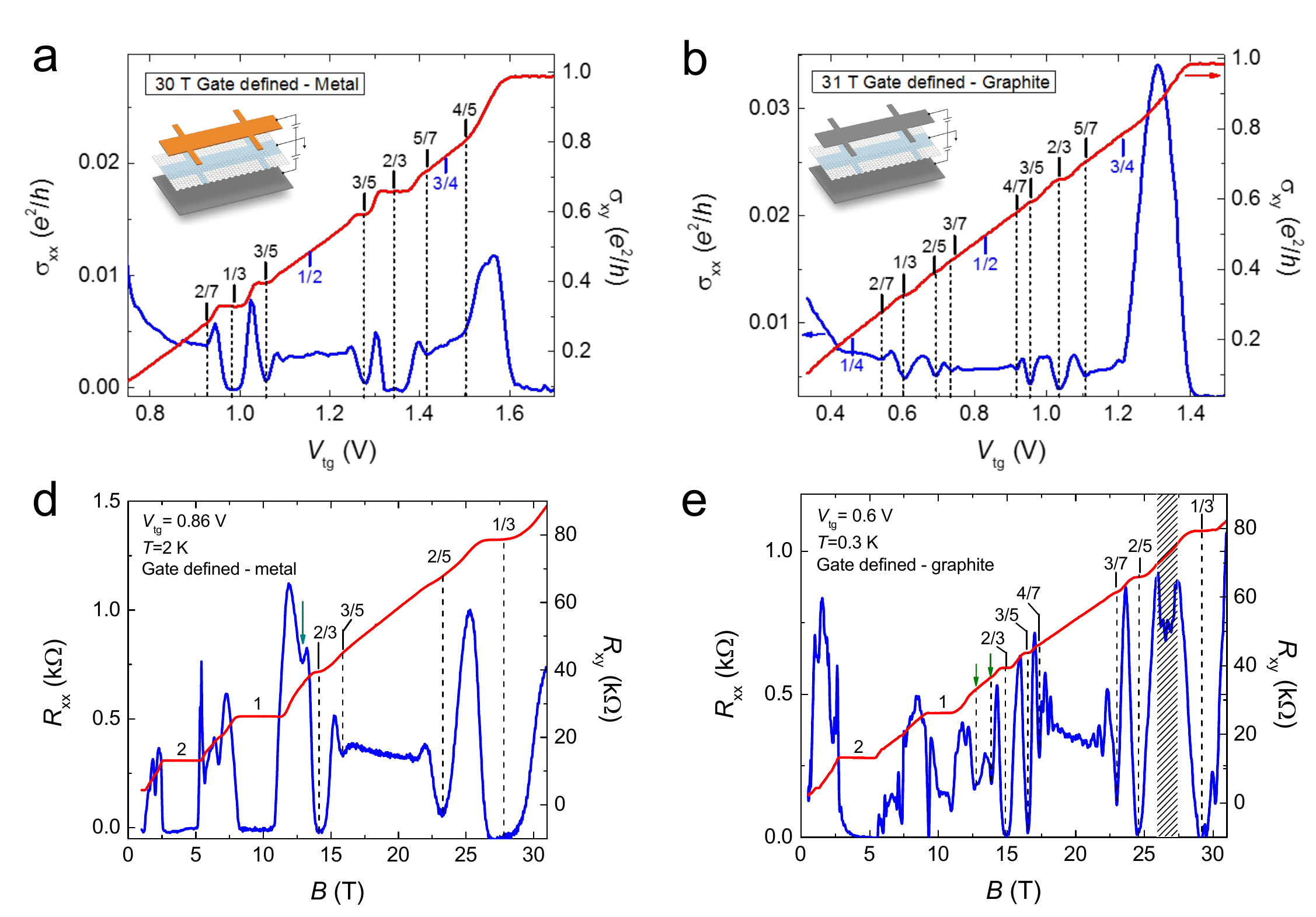}  
\caption{{\bf High field characterization}.  {\bf a} and {\bf b}, gate voltage sweep for the two devices at high magnetic field and low temperature.  {\bf c} and {\bf d}, magnetic field sweep for the same devices.}
\label{FieldCharac}
\end{figure}

In Fig. \ref{FieldCharac} we present some extra measurements of these devices which shows the enhanced  FQHE in the graphite gate-defined.

\section{Effects of the insulating state $\nu=0$}

\begin{figure}[h]
\centering
\includegraphics[scale=0.5]{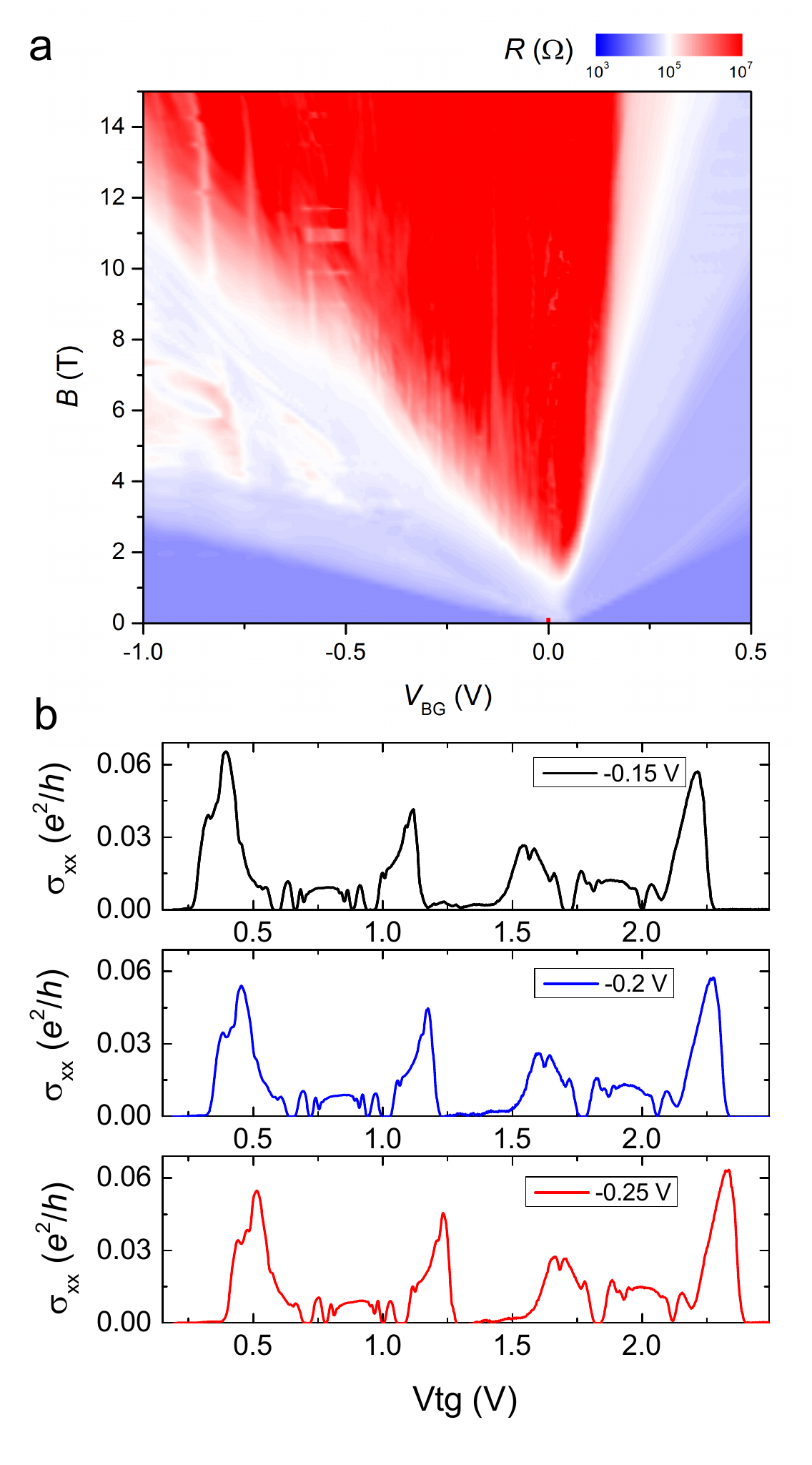}  
\caption{{\bf Insulating state}. {\bf a}, Two probes resistance of the uncovered section of the device as a function of bottom gate voltage and magnetic field, $T=0.3$ K. {\bf b}, longitudinal conductance of a graphite gate defined sample at 15 T and 0.3 K for different values of the bottom gate.}
\label{InsulatingState}
\end{figure}

The $\nu=0$ state is a exchange-enhanced energy gap where the bulk and edges of the sample are gapped. In Fig. \ref{InsulatingState} we present the magnetic field dependence of the insulating state at low temperature. This plot shows that the insulating state appear at rather low magnetic field and extend all the way to high magnetic fields. As we increase the magnetic field the voltage range of this gap is enlarged proving that our measurements are not jeopardize by a gap closing and it is not limited to a short range of bottom gate voltage. \\

In our measurements different values of the bottom gate voltage do not affect the FQH response.  We can see in Fig. \ref{InsulatingState}b that the FQH response of this graphite gate-defined sample is not affected by the voltage value of the bottom gate as long as it is in the insulating regime. 

\begin{figure*}[h]
\centering
\includegraphics[scale=0.45]{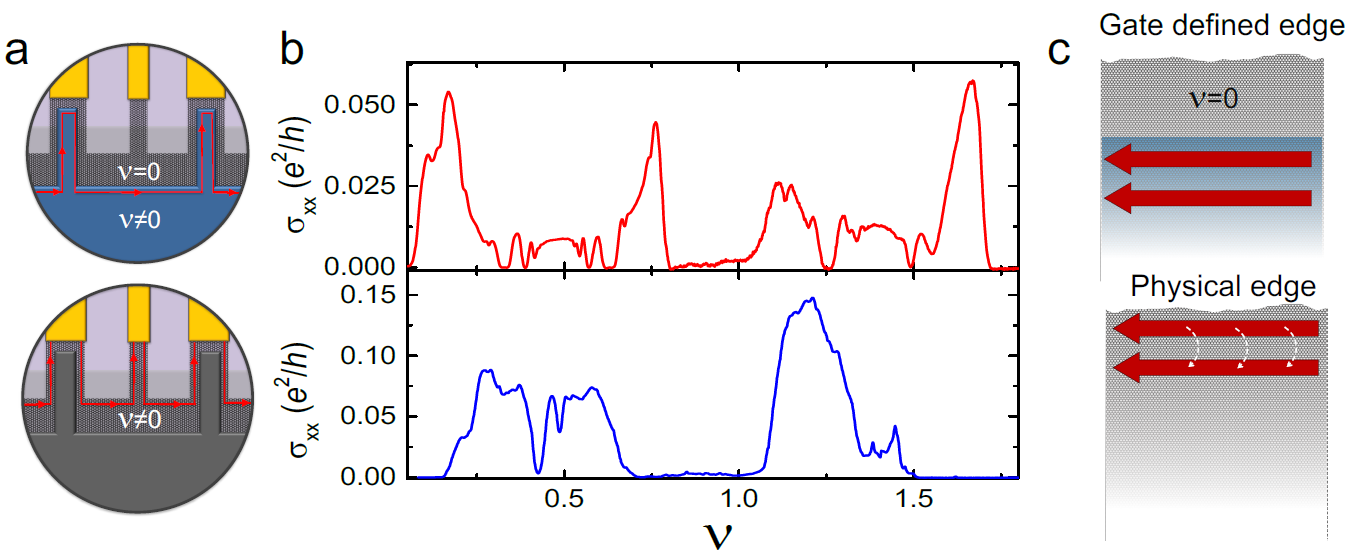}  
\caption{{\bf Gate defined and crystal edge}. {\bf a}, sketch of the edge states circulating around the electrostatically defined edges (top) and on the crystal edge (bottom)  {\bf b}, longitudinal conductance of a graphite gate defined sample at 15 T and 0.3 K for a gate top gate sweep (top) in the gate defined regime and a bottom gate sweep (bottom) with the top gate at 0 V. {\bf c}, cartoon of electrostatically defined edges (top) and physical edges (bottom).}
\label{InsulatingState2}
\end{figure*}

If the sample is not in the gate defined regime a mix of states occurs and the quantization of states is jeopardize. In the extreme case where the to gate is set to 0 V and the back gate changes the whole Fermi energy of the system the edge states will circulate on the crystal edge, Fig.  S\ref{InsulatingState2}a, we can see that the number of FQH states is highly reduced. In the case where the edge states circulate on the physical edge of the sample the in-homogeneous electrostatic landscape destroys the signature of the FQHE.

\begin{figure}[h]
\centering
\includegraphics[scale=0.7]{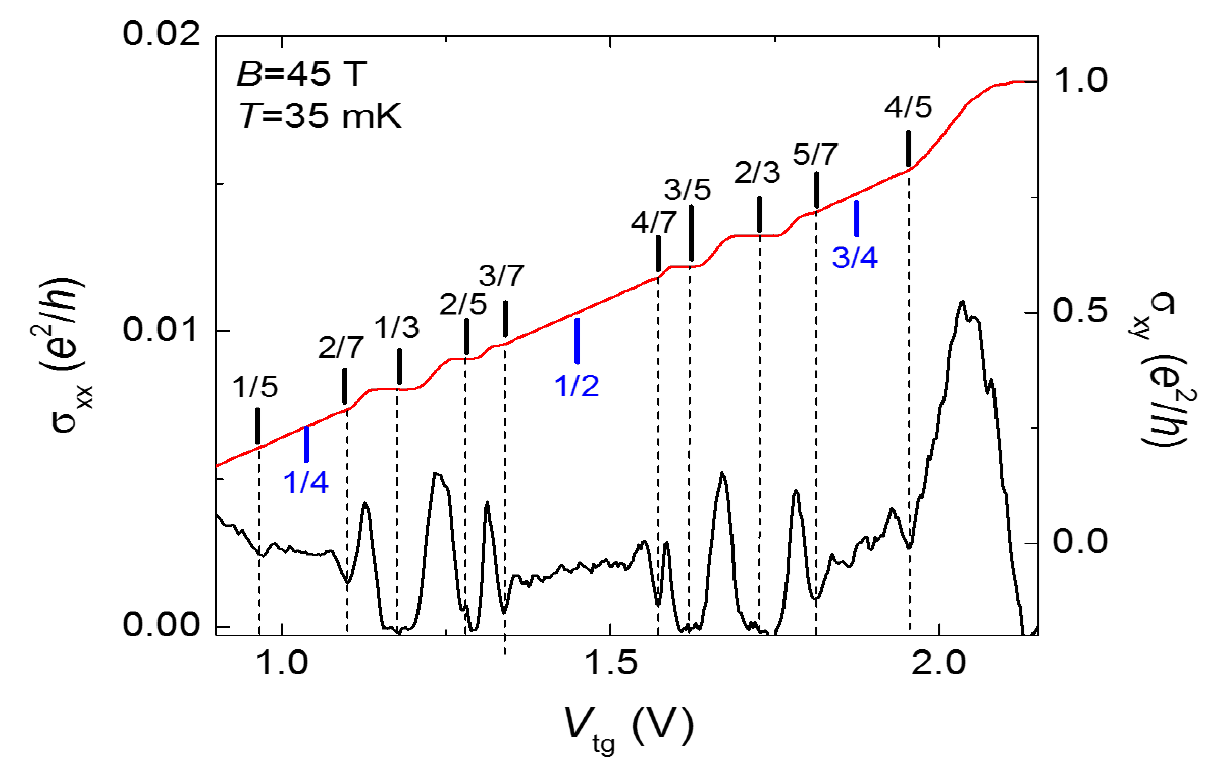}  
\caption{{\bf Four-flux state in a metal gate defined device at high magnetic field}. }
\label{s3}
\end{figure}

\begin{figure}[h]
\centering
\includegraphics[scale=0.7]{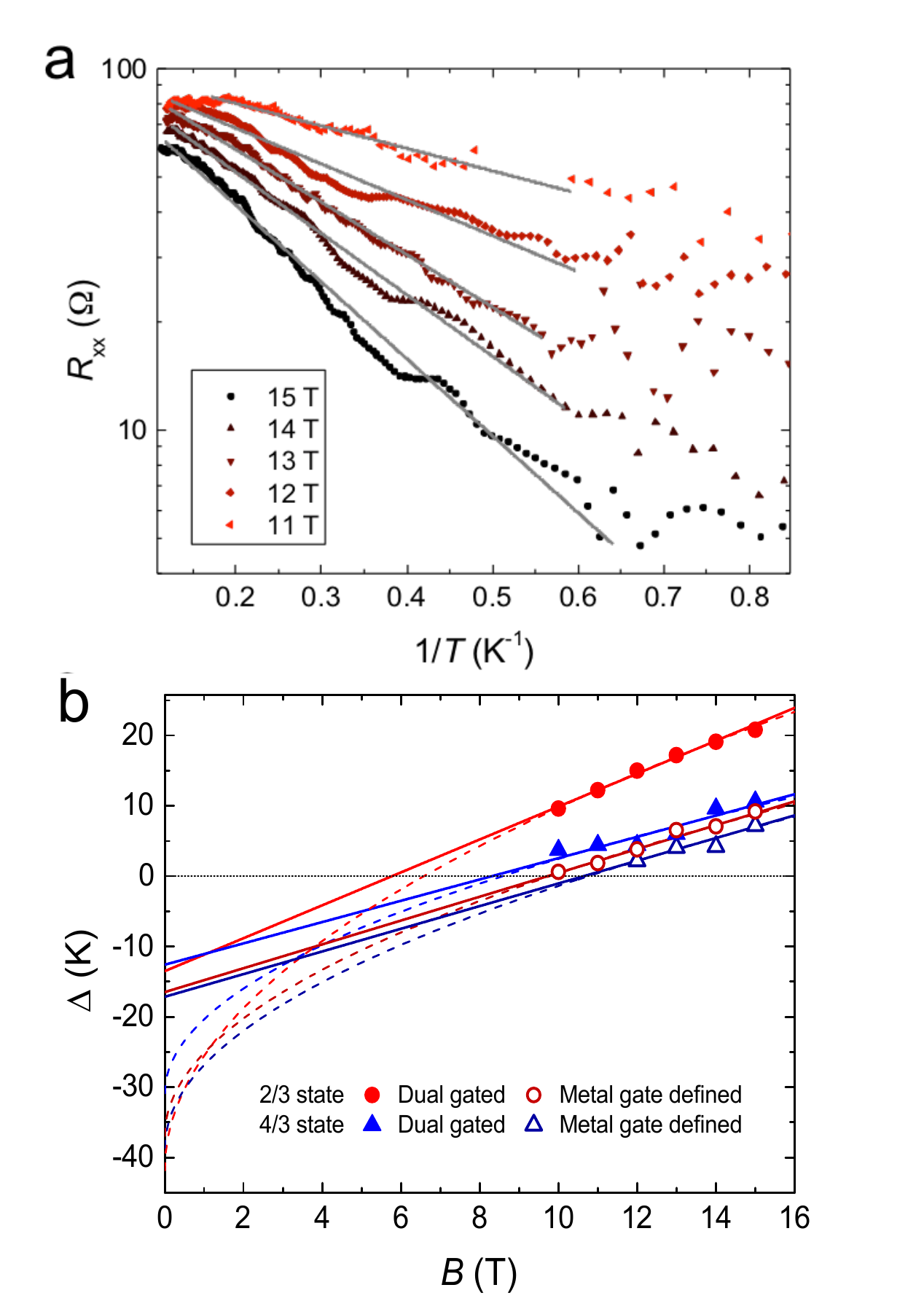}  
\caption{{\bf Magnetic field dependence of the energy gaps}. {\bf a}, Arhenious plot for different magnetic fields for the metal gate defined sample, 2/3 state. {\bf b}, Magnetic field dependence of the energy gaps for metal gate defined and dual gated for the 2/3 and 4/3 states. Dashed  and solid lines represent $\sqrt{B}$  and linear dependence, respectively. }
\label{s7}
\end{figure}

\begin{figure}[h]
\centering
\includegraphics[scale=0.25]{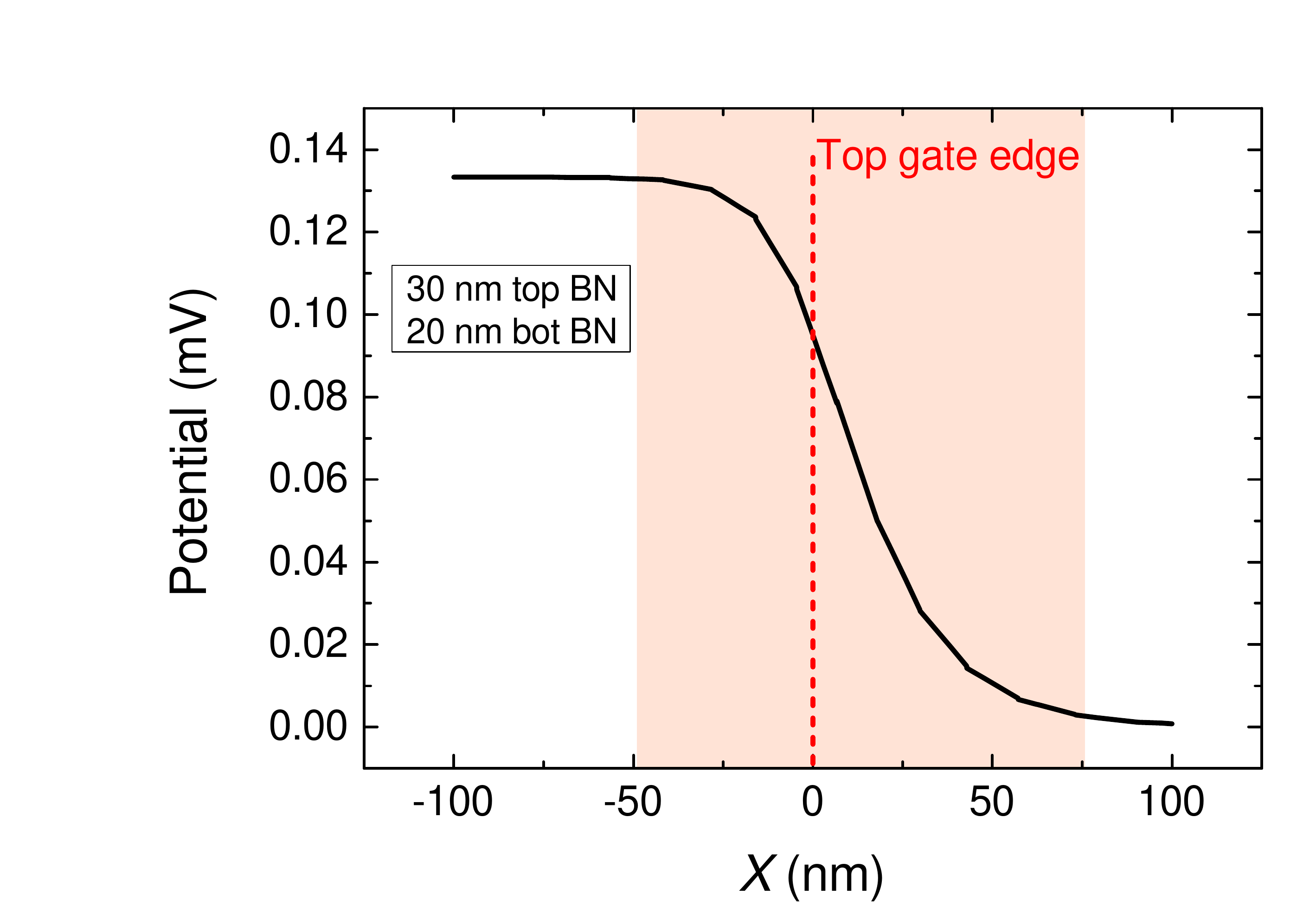}  
\caption{{\bf Electrostatic profile simulation.} numerical simulation using Comsol for a 30 nm top BN and a 20 nm bottom BN.}
\label{s9}
\end{figure}

\end{document}